# Ultrashort Pulse Detection and Response Time Analysis Using Plasma-wave Terahertz Field Effect Transistors

Yuhui Zhang and Michael S. Shur, *Life Fellow, IEEE*

*Abstract*—We report on the response characteristics of plasmonic terahertz field-effect transistors (TeraFETs) fed with femtosecond and picosecond pulses. Varying the pulse width ($t_{pw}$) from $10^{-15}$ s to $10^{-10}$ s under a constant input power condition revealed two distinctive pulse detection modes. In the short pulse mode ($t_{pw} \ll L/s$, where $L$ is the gated channel length, $s$ is the plasma velocity), the source-to-drain voltage response is a sharp pulse oscillatory decay preceded by a delay time on the order of $L/s$. The plasma wave travels along the channel like the shallow water wave with a relatively narrow wave package. In the long pulse mode ($t_{pw} > L/s$), the response profile has two oscillatory decay processes and the propagation of plasma wave is analogues to oscillating rod with one side fixed. The ultimate response time at the long pulse mode is significantly higher than that under the short pulse conditions. The detection conditions under the long pulse mode are close to the step response condition, and the response time conforms well to the analytical theory for the step function response. The simulated waveform agrees well with the measured pulse response. Our results show that the measurements of the pulse response enable the material parameter extraction from the pulse response data (including the effective mass, kinematic viscosity and momentum relaxation time).

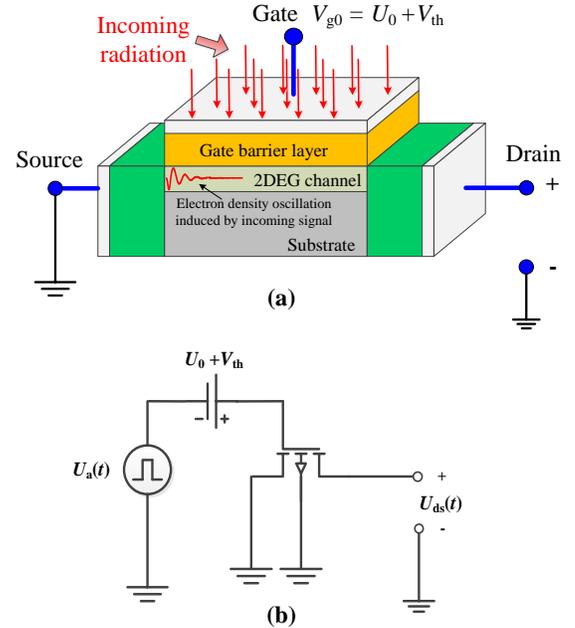

Fig. 1. (a) Schematic of a TeraFET under external radiation and (b) the equivalent biasing regime. $V_{th}$ represents the threshold voltage.

## I. INTRODUCTION

Plasma-wave terahertz field-effect transistors (TeraFETs) are recognized in recent decades as a promising candidate for THz applications [1-3]. TeraFETs have been developed in various material systems [4-8] and show the potential in such fields as THz detection [9], generation [10], imaging [11] and wireless communication [12]. To accelerate the commercialization of those THz applications, one of the key efforts is the detection and characterization of short pulses [9, 13].

The generation of high-power ultrashort pulses in the THz range is crucial for the ultra-fast time-domain spectroscopy [14-16]. To monitor the pulsed sources, high-quality sub-picosecond detectors are required. Schottky diodes [17-19] and quantum-engineered devices [20, 21] were utilized in ultrashort THz pulse detection and became the most commercialized devices in the relevant field. However, the operation of those devices requires strict condition (including, in some cases cryogenic cooling) and has limited operating ranges [15, 20, 21]. Plasma-wave THz detectors are not subject to those limitations [22, 23] and have the potential of reaching better performance as THz pulse detectors.

Fig. 1 shows the structure of a 2DEG TeraFET and the equivalent biasing circuit. A pulsed or continuous wave THz signal shining onto the TeraFET excites the oscillations in gate voltage, resulting in the electron density oscillations (plasma waves) in the TeraFET channel. These oscillations generate a voltage response signal $U_{ds}(t)$. We use this detected response for the THz excitation characterization.

Compared to more conventional THz detectors, plasmonic TeraFETs' advantages include tuneability, high responsivity, operation up to at least 6 THz, and demonstrated potential for THz sources [2, 24]. Besides, the response signals of plasmonic TeraFETs contain the information from the TeraFET materials

The authors are with the Department of Electrical, Computer and Systems Engineering, Rensselaer Polytechnic Institute, Troy, NY 12180 USA (e-mail: shurm@rpi.edu, zhangy79@rpi.edu).

TABLE I
MATERIALS PARAMETERS USED IN THE SIMULATION

| Material | $m/m_0$ | $\mu$ (77 K) | $\mu$ (300 K) | $\tau$ (77 K) | $\tau$ (300 K) | Reference |
|---|---|---|---|---|---|---|
| p-diamond | 0.74 | 35000 | 5300 | 14.75 | 2.23 | [8, 26, 27] |
| n-diamond | 0.36 | 50000 | 7300 | 10.25 | 1.50 | [8, 26, 28] |
| Si | 0.19 | 20000 | 1450 | 2.16 | 0.157 | [29-33] |
| AlGaN/GaN | 0.24 | 31691 | 2000 | 4.33 | 0.273 | [30, 34, 35] |
| AlGaAs/InGaAs | 0.041 | 35000 | 12000 | 0.817 | 0.280 | [9, 36, 37] |

*Note: $m$, $m_0$ are the effective mass of carriers and the free electron mass, respectively. $\tau$ (in ps) and $\mu$ (in cm$^2 \cdot$V$^{-1}$s$^{-1}$) are the momentum relaxation time and the mobility of carriers, respectively.*

suitable for materials characterization and 2DEG parameter extraction [9]. Plasmonic TeraFETs can also be used as broadband THz pulse emitters and mixers [10, 25].

In this work, we explore the potential of plasmonic TeraFETs as ultrashort pulse detectors. By focusing on the previously unknown characteristics such as pulse width dependence of response time and material parameter extraction, we aim to enrich the basic knowledge and exploit more applications of TeraFET pulse detectors.

## II. BASIC EQUATIONS

A one-dimensional hydrodynamic model is employed in this work to explore the excitation of plasma waves and response features in gated plasmonic TeraFETs. 5 material systems, namely p-diamond, n-diamond Si, GaN, and InGaAs, are implemented in this study. The corresponding materials parameters are listed in Table I [8, 9, 26-37].

The governing hydrodynamic model equations are [5, 38]

$$\frac{\partial n}{\partial t} + \nabla \cdot (n\boldsymbol{u}) = 0 \quad (1)$$

$$\frac{\partial \boldsymbol{u}}{\partial t} + (\boldsymbol{u}\cdot\nabla)\boldsymbol{u} + \frac{e}{m}\nabla U + \frac{\boldsymbol{u}}{\tau} - v\nabla^2\boldsymbol{u} = 0 \quad (2)$$

$$\frac{\partial \theta}{\partial t} + \nabla\cdot(\theta\boldsymbol{u}) - \frac{\chi}{C_v}\nabla^2\theta - \frac{mv}{2C_v}\left(\frac{\partial u_i}{\partial x_j} + \frac{\partial u_j}{\partial x_i} - \delta_{ij}\frac{\partial u_k}{\partial x_k}\right)^2 \quad (3)$$
$$= \frac{1}{C_v}\left(\frac{\partial W}{\partial t}\right)_c + \frac{m\boldsymbol{u}^2}{C_v\tau}$$

Here, $n$ is the carrier density, $\boldsymbol{u}$ represents the hydrodynamic velocity. $U$ is the gate-to-channel potential defined as $U = U_0 - U_{ch}$, where $U_0$ and $U_{ch}$ are the gate bias above threshold and the channel potential, respectively. In the simulation, we use gradual channel approximation ($CU = en$, where $C$ is the barrier layer capacitance fixed at 0.56 μF/cm$^2$ [5]) to associate $U$ and $n$. Besides, $\tau = \mu m/e$ is the momentum relaxation time of carriers, where $m$ and $\mu$ represents the effective mass and the mobility, respectively.

In the energy equation (3), $\theta = k_BT$ represents the carrier temperature in eV, $\chi$ is a normalized heat conductivity defined as $\chi = \kappa/n$ ($\kappa$ is the heat conductivity); $C_v$ is the thermal capacitance and $C_v = (\partial\Sigma/\partial\theta)_n$ [9, 38], where $\Sigma = \theta*F_1(\xi)/F_0(\xi)$ is the average internal energy, $F_k(\xi)$ is the Fermi integral, $\xi = \ln(\exp(E_F/k_BT)-1)$ is the chemical potential, $E_F = k_BT_F = \pi\hbar^2n/m$ is the Fermi energy. $W$ is the total energy, and $(\partial W/\partial t)_c$ represents the collision term of $\partial W/\partial t$, which can be expressed by $(\partial W/\partial t)_c = (\partial\Sigma/\partial t)_c - m\boldsymbol{u}^2/\tau$. Furthermore, $v$ is the viscosity of the 2DEG/2DHG, which is related to the carrier density and temperature. When the temperature $T$ is much lower than the Fermi temperature $T_F$, $\chi$ and $v$ are given by [5, 9, 38]:

$$v(T) = \frac{2\hbar}{\pi m}\frac{T_F^2}{T^2}\frac{1}{\ln(2T_F/T)} \quad (T \leq T_F) \quad (4)$$

$$\chi(T) = \frac{4\pi\hbar}{3m}\frac{T_F}{T}\frac{1}{\ln(2T_F/T)} \quad (T \leq T_F) \quad (5)$$

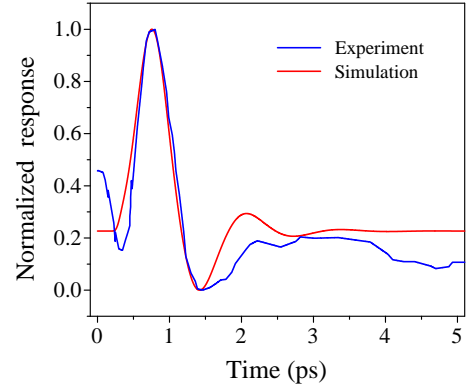

Fig. 2. Comparison between the normalized voltage response obtained from our simulation and the experimental result by Shur *et al*. [43]. The parameters used in the simulation are: $L = 220$ nm, Gaussian pulse (center: 440 fs, standard deviation: 200 fs), $U_0 = -0.1$ V, $v = 26$ cm$^2$/s, $m = 0.041m_0$, $\mu = 12000$ cm$^2\cdot$V$^{-1}$s$^{-1}$. The normalization is performed by $U_{norm} = (U_{r0} - U_{min})/(U_{max} - U_{min})$, where $U_{norm}$ and $U_{r0}$ represents the normalized and original voltage response, respectively; $U_{max}$ and $U_{min}$ are the maximum and minimum of $U_{r0}$, respectively

Generally, $T < T_F$ holds for relatively large gate bias (e.g. $U_0 > 2.3$ V for p-diamond TeraFETs), at which the viscosity and heat conductivity could be large. For relatively small $U_0$, $T > T_F$, and we use $v(T = T_F)$ and $\chi(T = T_F)$, which are constant values [5].

The boundary condition that we use are an open drain condition [5, 23]: $U(0, t) = U_0 + U_a(t)$ and $J(L, t) = 0$, where $U_a(t)$ represents the pulsed small signal voltage induced by the incoming radiation, $J$ is the current flux density. This boundary condition agrees with the typical biasing regime shown in Fig. 1(b). It is worth noting that $U_a(t)$ is used to model the excitation of gate-to-channel voltage by the external radiation, and such modeling has been proved to be effective in previous studies [5, 22, 23, 25, 39]. Besides, in this work we only consider the detection of single-pulse signals, i.e. $U_a(t)$ contains only one pulse without repetition. Also, all the pulses considered are square-shaped. Thus $U_a(t)$ can be expressed by $U_a(t) = U_{am}(u(t)-u(t-t_{pw}))$, where $U_{am}$ is the amplitude of the pulse voltage, $u(t)$ is the unit step function, $t_{pw}$ is the pulse width.

The above hydrodynamic model is valid when $1/\tau_{ee} \gg 1/\tau$ [5, 9, 38, 40, 41], where $\tau_{ee}$ is the electron–electron scattering rate, $\tau$ represents the momentum relaxation time. We have checked, by calculating the value of $\tau_{ee}$ [42], that this condition is met in this study. More detailed introductions of the hydrodynamic model can be found in [5], [38] and [41]. The results obtained from the hydrodynamic model were compared to experimental measurements in both pulse detection [43] and continuous wave detection mode [5, 44-48] in previous studies, and some good agreement between modeling and experimental data were obtained. Here we also compare our simulation result of voltage response with previous experimental data, as illustrated in Fig. 2. The experimental data come from Shur *et al.* Frontiers in Electronics, March 2017, 21-34 (2017) [43]. As seen, the simulation curve shows a fair qualitative (i.e. the evolution trends, peak moments *etc.*) agreement with the experimental data, which proved the effectiveness of this model.

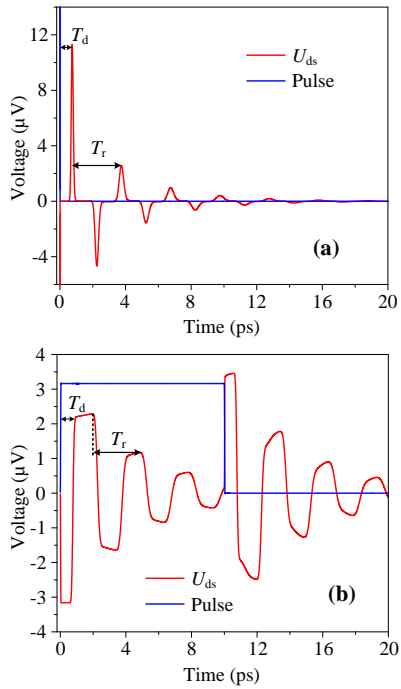

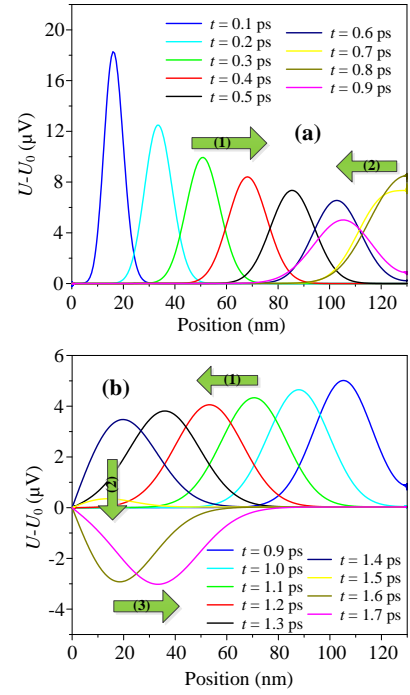

Fig. 3. Temporal response profile for p-diamond TeraFET at (a) $t_{pw} = 10^{-14}$ s, (b) $t_{pw} = 10^{-11}$ s. The DC gate bias $U_0$ is 0.1 V, the channel length $L$ is 130 nm, and the operating temperature $T$ is 300 K. Under above conditions, $L/s \approx 0.8$ ps.

Fig. 4. Spatial profiles of voltage ($U - U_0$) at different moments: (a) 0.1-0.9 ps, (b) 0.9-1.7 ps, for p-diamond TeraFET at $t_{pw} = 10^{-14}$ s. The green arrows and numbers within show the propagation trend of wave fronts and the sequence of steps, respectively.

## III. RESULTS AND DISCUSSION

### A. Short pulse versus long pulse

By varying the pulse width of incoming single-pulse signal from $10^{-15}$ s to $10^{-10}$ s under a constant input power condition (i.e. $U_{am}^2 t_{pw}/R_{ch}$ = constant = $10^{-22}$ V$^2 \cdot$s/$R_{ch}$, where $R_{ch}$ (in Ω) is the channel resistance), the effect of pulse width on the response characteristics of TeraFETs is investigated. In the simulation, two distinctive response modes are observed: (1) the short pulse mode, in which $t_{pw} \ll L/s$, and the long pulse mode with $t_{pw} > L/s$. We first take a look at the temporal profiles of voltage response $U_{ds}$ under the short and long pulse conditions, taking $t_{pw} = 10^{-14}$ s and $t_{pw} = 10^{-11}$ s cases as examples. The results are presented in Fig. 3. As shown in Fig. 3(a), some sharp voltage response peaks are observed, and those peaks are generated after the pulsating period due to a short pulse width of input signal. A delay time $T_d$ on the order of $L/s$ is observed between the pulse and the first response peak, which is a signature of plasmonic ballistic transport in collision-dominated devices [9]. Besides, the amplitude of voltage peak decays with time. Those features indicate that the device is operated at underdamped plasmonic resonant regime [9, 23, 49]. Fig. 3(b) presents the temporal response waveform at the long pulse mode. Unlike the short pulse case, two oscillatory decay processes are observed in the voltage response waveform. One decay process is within the pulsation period, the other is initiated right after the incoming pulse ends. This phenomenon has never been reported in previous studies. Furthermore, compared to the short pulse case, the voltage response pulses in Fig. 3(b) are much wider, and the shape of pulses resembles the square pulse.

To better understand the mechanisms underlying two detection modes, we trace the voltage distribution profile along the channel at several key moments, as shown in Fig. 4 and Fig. 5. Fig. 4 illustrates the spatial voltage ($U - U_0$) profile when $t_{pw} = 10^{-14}$ s. As seen, a voltage peak is generated near the source side ($x = 0$) at $t = 0.1$ ps, then it moves rightwards towards the drain ($x = 130$ nm). The movement of voltage wave is a reflection of the propagation of plasma wave in the channel, with a speed close to plasma velocity $s \approx \sqrt{(eU_0/m)}$. When the wave precursor hits the boundary, it gets reflected and changes direction like shallow water [22], as shown in Fig. 4(a) and Fig. 4(b). When reaching the source side, the polarity of the wave reverses due to the fixed source voltage, as shown in Fig. 4(b). Those features clearly demonstrate that the device is operating at plasmonic resonant mode.

Fig 5 illustrates the propagation characteristics of plasma wave for a long pulse ($t_{pw} = 10^{-11}$ s). As presented in Fig. 5(a), the voltage ramps up from source to the drain in the beginning phase. When the wave front reaches the drain, the voltage level there continues to ramp up, and then the voltage rise spreads leftwards and creates a reversed wave propagation process. When the reverse wave hits the source, it reflects and propagates back. As the propagation continues, the energy of the wave gradually dissipates due to the friction and viscosity, and the amplitude of the voltage oscillations attenuates with time. This dynamics of voltage distribution is more like a waving rod with one side fixed, which is very different from the short pulse case shown in Fig. 4. Fig. 5(b) illustrates the spatial voltage profiles at some moments of time right after the pulsating phase. One can see that the voltage at the source side falls due to the quenching of external excitation. This leads to the destabilization of the system and consequently initiates another wave propagation process. Thus, two oscillatory decay processes are observed in Fig. 3(b).

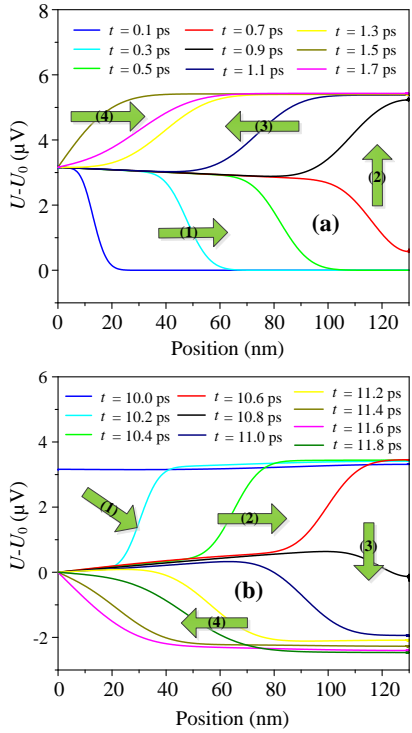

Fig. 5. Spatial profiles of voltage ($U - U_0$) for time periods: (a) 0.1-1.7 ps, (b) 10-11.8 ps, for p-diamond TeraFET at $t_{pw} = 10^{-11}$ s. The green arrows and numbers within show the propagation trend of wave fronts and the sequence of steps, respectively.

The two distinctive wave propagation mechanisms determine the ultimate response time, which is a very important parameter for the pulse detection. For weak incoming signals, the analytical response characteristics are evaluated by solving the linearized hydrodynamic equations and getting the solution in the form of $\sum_{(n)} A_n \exp(\sigma_n t) f_n(x)$, where $\sigma_n$ is the *n*-order attenuation factor [9]. The analytical response time $\tau_r$ can be approximated by the first order solution, i.e. $\tau_r = \mathrm{Re}\,[1/|\sigma_1|]$. To calculate the response time from the simulation data, we could fit the temporal voltage response into a periodic oscillatory decay $\exp(-t/\tau_r)\cdot\cos(2\pi t/T_r+\varphi)$, where $T_r$ is the period of oscillation ($\sim 2L/s$). However, at a low viscosity, the higher order modes become non-negligible and such fitting fails [5, 9]. To address this problem, we extract the peak points in the response curve and then do the exponential decay fitting to get the response time.

Fig 6 presents the variation of response time with the pulse width for the TeraFETs fabricated using 5 different materials systems. As seen, for all the TeraFETs the general evolution trends of $\tau_r$ are identical. When $t_{pw} \ll L/s$, the response time is relatively short and keeps almost unvaried with the increase of $t_{pw}$. As $t_{pw}$ approaches $L/s$, the response time experiences a sharp increase and then stabilizes as $t_{pw}$ rises beyond $L/s$. This indicates that the unique wave propagation under the long pulse mode leads to a significant increase of response time compared to short pulse cases.

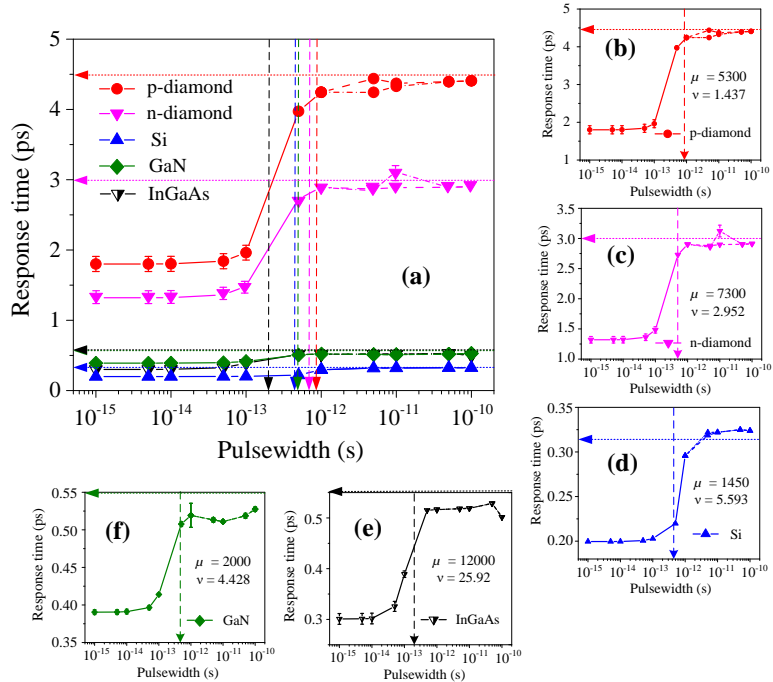

Fig. 6. Response time vs pulse width (a) for TeraFETs of 5 materials under $U_0 = 0.1$ V, $T = 300$ K, $L = 130$ nm. The dashed lines and dash-dot lines illustrate the response time variation for the first and second decay process, respectively. (b)-(f) show the separated response profiles for p-diamond, n-diamond, Si, InGaAs and GaN TeraFET, respectively. The horizonal dotted arrow lines and vertical dashed arrow lines show the values of $2\tau$ and $L/s$ for different materials (color-coded), respectively. The units of $\mu$ and $\nu$ are cm$^2\cdot$V$^{-1}$s$^{-1}$ and cm$^2\cdot$s$^{-1}$, respectively.

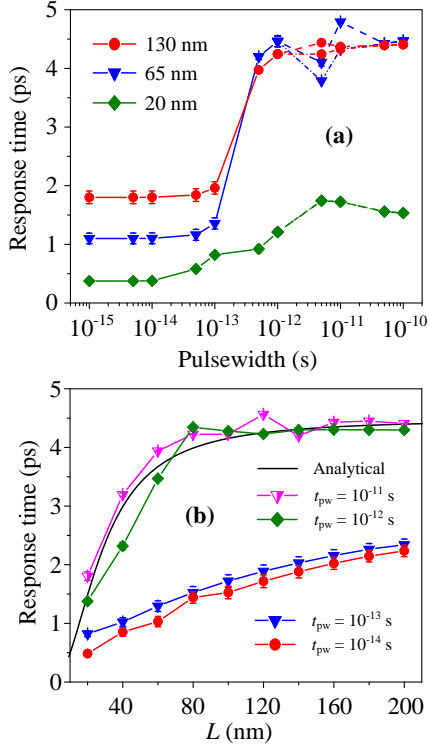

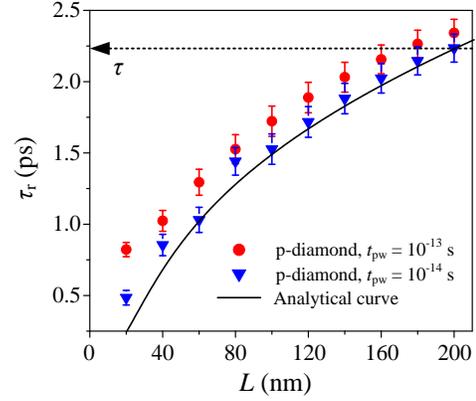

Fig. 8. Comparison between the simulated and analytical response time (equation (8)) as functions of $L$ under the short pulse mode. The dotted arrow line illustrates the value of $\tau$.

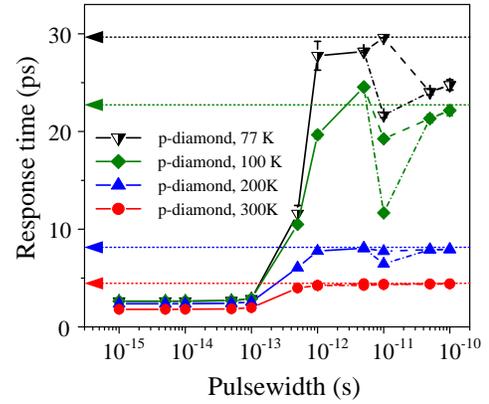

Fig. 7. (a) Response time vs pulse width under 3 different channel lengths. (b) Response time vs channel length under 4 different pulse width values for p-diamond TeraFETs. $U_0 = 0.1$ V, $T = 300$ K

### B. Feature size and temperature dependence

To further evaluate the pulse detection performances of TeraFETs, we study the feature size (channel length) and temperature dependence of response time. Fig. 7 presents the channel length dependence of response time, taking p-diamond TeraFET as the example. As shown in Fig. 7(a), as $L$ increases from 20 nm to 130 nm, the response time at $t_{pw} \ll L/s$ increases. For $t_{pw} > L/s$ cases, a dramatic increase is observed when $L$ rises from 20 nm to 65 nm. However, as $L$ increases further, the response time is saturated and no longer vary with $L$. To better understand such evolution trend, Fig. 7(b) plots the response time as a function of $L$ under 4 different values of pulse width. The analytical curve is also included in this figure. As mentioned earlier, the analytical solution is obtain by solving the linearized hydrodynamic equations in the form of $U(x, t) = \sum_{(n)} A_n \exp(\sigma_n t) f_n(x)$, and $\tau_r \approx 1/\text{Re}(|\sigma_1^+|)$ (first-order approximation). The value of $\sigma_n$ is given by [5, 9]

$$\sigma_n^\pm = \frac{1}{2}\left[-\left(\frac{1}{\tau}+\frac{\pi^2 \nu n^2}{4L^2}\right)\pm\sqrt{\left(\frac{1}{\tau}+\frac{\pi^2 \nu n^2}{4L^2}\right)^2 - \frac{\pi^2 s^2 n^2}{L^2}}\right] \quad (6)$$

$n = 1, 3, 5, ...$

As shown in Fig 7(b), for short pulse cases ($t_{pw} = 10^{-14}$ s and $10^{-13}$ s.), the response time rises with increasing $L$, but the values deviate from the analytical prediction. For long pulse cases ($t_{pw} = 10^{-12}$ s and $10^{-11}$ s.), a sharp increase in response time at around $L = 20$-60 nm is observed, and the simulation values agree with the analytical curve. This result is expected, since the analytical solution is obtained by setting an arbitrary initial condition and solving its transient response of linearized hydrodynamic equations, in which a quasi-steady-state condition is assumed [5]. In other word, the analytical theory is targeted on the step function input. For TeraFETs operated at a long pulse mode, the condition is close to the step function input during each decay process. Therefore, the calculated response times generally conform to the analytical values. For short pulse cases, however, the external excitation is extinguished before the initial plasma wave reaches the drain. Therefore, the voltage profile along the channel does not completely follow the $U(x, t) = \sum_{(n)} A_n \exp(\sigma_n t) f_n(x)$ form. This explains why the simulation curves deviate from the analytical curve for the short pulses. In addition, since the plasma waves in the short pulse mode appears to be less confined in the channel compared to those in the long pulse cases, they are more subject to the electronic fluid friction (i.e. scattering) or viscosity and thus decay more quickly, as shown in Fig. 3. This might be the reason why the response time at short pulse mode is much smaller compared to that of the large pulse mode.

Now we briefly discuss the variation trends of response time shown in Fig. 7. According to the analytical theory, under the high mobility condition ($\tau \gg L/s$), the expression of $\tau_r$ can be approximated by

$$\tau_r \approx \frac{2\tau}{1+\pi^2\nu\tau/4L^2} \quad (\tau \gg L/s,\ t_{pw} > L/s) \quad (7)$$

Fig. 9. Response time vs pulse width for p-diamond TeraFET under 4 different temperatures. $U_0 = 0.1$ V, $L = 130$ nm. The dashed lines and dash-dot lines illustrate the response time variation for the first and second decay process, respectively. The horizonal dotted arrow lines show the values of $2\tau$ at different temperatures (color-coded).

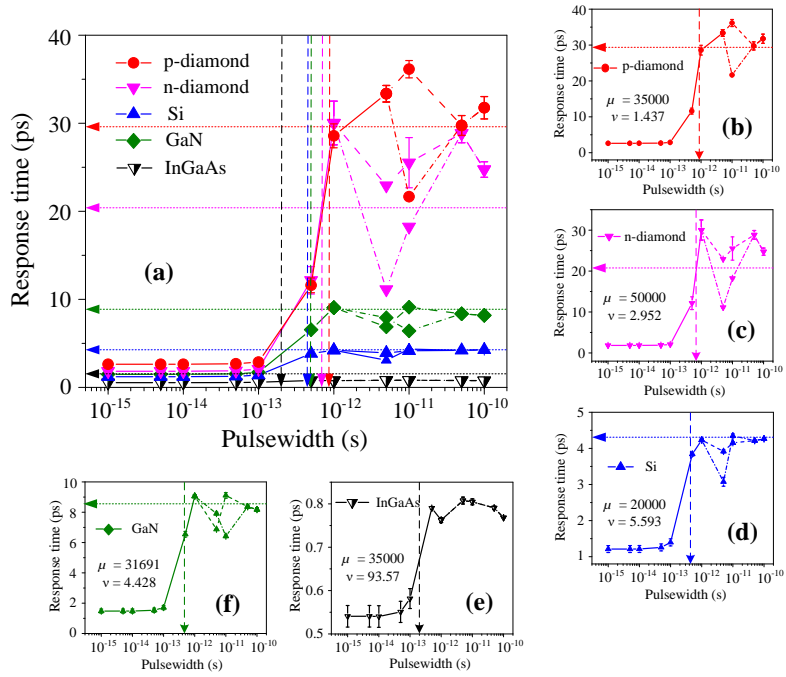

Fig. 10. Response time vs pulse width (a) for TeraFETs of 5 materials under $U_0 = 0.1$ V, $T = 77$ K, $L = 130$ nm. The dashed lines and dash-dot lines illustrate the response time variation for the first and second decay process, respectively. (b)-(f) show the separated response profiles for p-diamond, n-diamond, Si, InGaAs and GaN TeraFET, respectively. The horizonal dotted arrow lines and vertical dashed arrow lines show the values of $2\tau$ and $L/s$ for different materials (color-coded), respectively. The units of $\mu$ and $\nu$ are cm$^2\cdot$V$^{-1}$s$^{-1}$ and cm$^2\cdot$s$^{-1}$, respectively.

When $L$ is relatively small, the term $\pi^2\nu\tau/4L^2$ decreases rapidly with the increase of $L$, thus a sharp increase in $\tau_r$ is observed. As $L$ increases further, $\pi^2\nu\tau/4L^2$ drops to very small values, so that $\tau_r \Rightarrow 2\tau$ and $\tau_r$ becomes less $L$ dependent. This explains why $\tau_r$ tends to saturate for the high $L$ values for long pulse cases.

As regards the short pulse cases, it is noticed from Fig. 7(b) that the values of $\tau_r$ for $t_{pw} = 10^{-14}$ s and $t_{pw} = 10^{-13}$ s cases are still on the order of momentum relaxation time, and have a very weak $t_{pw}$ dependence. Also, $\tau_r$ appears to follow a $L^{0.5}$ law. Therefore, we propose a new semi-empirical expression of $\tau_r$:

$$\tau_r \approx \frac{2\tau}{1+\pi^2\nu\tau/4L^2}\sqrt{\frac{L}{2s\tau}} \quad (\tau \gg L/s,\ t_{pw} \ll L/s) \quad (8)$$

The term $(L/2s\tau)^{0.5}$ accounts for the plasmonic acceleration in the short pulse mode. The comparison between the results of equation (8) and the simulation data is given in Fig. 8. As seen, a fair agreement is achieved between the simulation points and the analytical curve.

In addition to the gate length dependence, it is also important to investigate the temperature dependence of $\tau_r$. Fig. 9 presents $\tau_r$ as a function of $t_{pw}$ under 4 different temperatures. The mobilities for $T = 300$ K, 200 K, 100 K and 77 K cases are set as 5300 cm$^2$/s, 9700 cm$^2$/s, 27000 cm$^2$/s, and 35000 cm$^2$/s, respectively [8, 26]. As seen from Fig. 9, with the decrease of $T$, the response times in both short pulse and large pulse cases rise due to the increase of the momentum relaxation time. Moreover, the increase of response time with a temperature decrease in the long pulse regime is more significant compared to that for the short pulse regime. This confirms to our proposed equation (8) (i.e. the additional term $(L/2s\tau)^{0.5}$ suggests that $\tau_r \propto \tau^{0.5}$ rather than $\tau_r \propto \tau$ in the short pulse mode). The result also indicates that under the short pulse mode, the TeraFET devices have a better temperature stability in terms of response characteristics, and thus could be more suitable for multi-temperature pulse detection. For long pulses, the variation of $\tau_r$ with $T$ is greater. This is related to the increase of $\tau$ at low temperatures and to the deviation of $\tau_r$ from the $2\tau$ limit [5]. As $\tau_r$ rises, the plasma waves are less attenuated within the pulse period affecting the initial condition for the second decay process. Therefore, the discrepancy in $\tau_r$ between the first and second decay process is enlarged as seen in Fig. 9.

To compare the detection performance of all types of TeraFETs at cryogenic temperature, Fig 10 illustrates the variation of response time with $t_{pw}$ for the TeraFETs fabricated using 5 different materials systems at 77 K. As seen, the general variation trend of response time is the same as that at 300 K, but the level of $\tau_r$ increases due to the increase of mobility and $\tau$. At 77 K, the InGaAs TeraFET has the lowest response time in both short pulse and long pulse mode, and the values of $\tau_r$ at long pulse mode are much lower than $2\tau$, as shown in Fig. 10(e). This is because the increase of mobility in InGaAs TeraFET from 300 K to 77 K is less prominent compared to other TeraFETs, and a high viscosity significantly attenuates the response time. It also suggests that InGaAs TeraFETs may be most suitable as fast response detectors at low temperatures. Besides, p-diamond and n-diamond TeraFETs still have relatively large response times compared with other TeraFETs, and the variation of $\tau_r$ between the first and second decay process at long pulse mode is enlarged compared to 300 K case shown in Fig. 6. This phenomenon is consistent with the trend shown in Fig. 9.

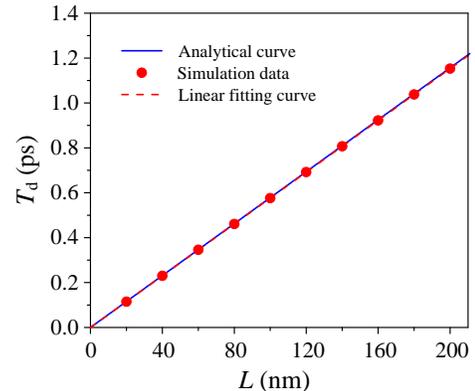

Fig. 11. Delay time $T_d$ as a function of channel length in p-diamond TeraFETs at $t_{pw} = 10^{-14}$ s. $U_0 = 0.1$ V, $T = 300$ K.

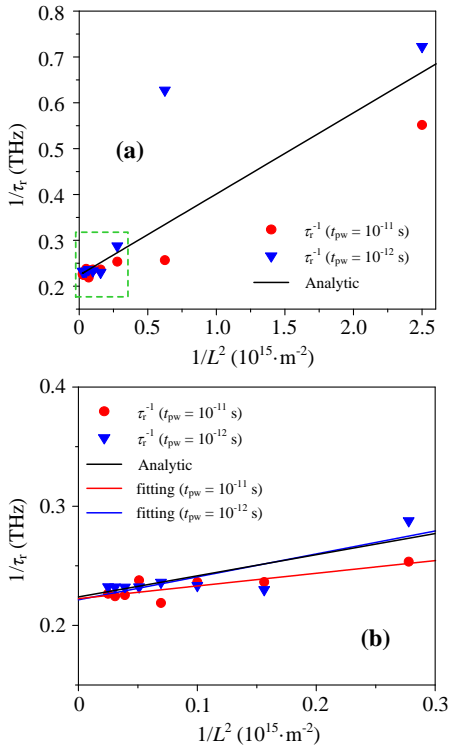

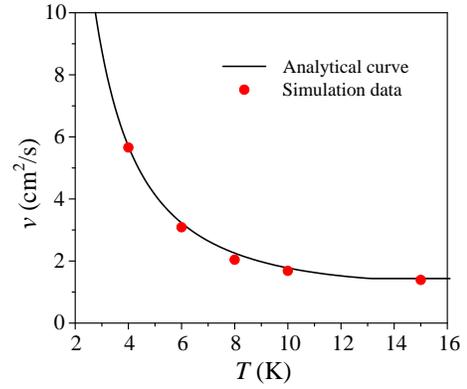

Fig. 12. (a) Simulation data and analytical curve of $\tau_r^{-1}$ vs $L^{-2}$ and (b) the zoom-in plot of the green dashed box in (a) (relatively high $L$).

Fig. 13. Temperature dependence of the viscosity for p-diamond TeraFET. The simulation scatters are obtained by the $\tau_r^{-1}$ - $L^{-2}$ fitting method mentioned above, while the analytical curve follows equation (4).

### C. Parameter extraction from the voltage response data

Our results show that the response of a TeraFET to variable width pulses should allow for the materials parameter extraction.

1) Extraction of effective mass $m$.

In order to get the effective mass $m$, the delay time $T_d$ is used. Under the resonant regime, the delay time $T_d$ is in fact the time for plasma waves to travel from source to drain, so $T_d$ should be on the order of $L/s$. The approximate expression of $s$ is given by $s \approx s_e = \sqrt{eU_0/m}$. For a more accurate calculation, one also need to consider the contribution of pressure gradient (acoustic velocity $s_{ac}$), $s_{ac} = (k_B T/m)^{0.5}$ (for non-degenerate case) so that $s = (s_e^2 + s_{ac}^2)^{0.5}$ [9]. Then we obtain the expression for $T_d$:

$$T_d \approx \frac{L}{s} = L\sqrt{\frac{m}{eU_0 + k_B T}} \quad (9)$$

Therefore, the effective mass can be obtained by plotting the $T_d$ vs $L$ fitting curve and extracting the slope of curve. Fig. 11 presents both simulation and analytical data of $T_d$ as a function of $L$ for p-diamond TeraFETs at $t_{pw} = 10^{-14}$ s. As seen, the simulation results show a very good agreement with the analytical prediction. The linear fitting of the simulation points gives a slope $k = 5.77 \times 10^{-6}$ s/m, yielding $m = (eU_0 + k_B T)k^2 = 6.711 \times 10^{-31}$ kg $= 0.737 m_0$, which is very close to the value used in the simulation ($m = 0.74 m_0$). For other $t_{pw}$ values and other materials under the short pulse mode, similar accuracy is obtained using equation (9) to extract the effective mass.

2) Extraction of $v$ and $\tau$:

The kinematic viscosity $v$ and the momentum relaxation time $\tau$ are two key parameters that determine the performance of the plasmonic devices. The momentum relaxation time is directly associated with the decay of plasma waves and thus determines the response performance of TeraFETs in both continuous wave detection and pulse detection mode [5, 9, 23, 48]. The viscosity also has a significant impact on the amplitude of the response and the decay speed of plasma waves. Also, the value of $v$ determines the proportion of high-order harmonics of the response signal [9]. Rewriting Eq. (7), we get

$$\tau_r \approx \frac{2\tau}{1+\pi^2 v\tau/4L^2} \Rightarrow \frac{1}{\tau_r} = \frac{1}{2\tau} + \frac{\pi^2 v}{8}\frac{1}{L^2} \quad (10)$$

By plotting $\tau_r^{-1}$ vs $L^{-2}$ (for the long pulse data), we could get $\tau$ from the intercept and $v$ from the slope.

Fig. 12 presents the simulation and analytical curves of $\tau_r^{-1}$ with varying $L^{-2}$. It can be seen that the simulation data generally agree with the analytical curve. For $t_{pw} = 10^{-11}$ s case, the intercept and slope of fitting curve are $2.209 \times 10^{11}$ s$^{-1}$ and $1.328 \times 10^{-4}$ m$^2$/s, respectively, correspond to $\tau = 2.263$ ps and $v = 1.076$ cm$^2$/s. For $t_{pw} = 10^{-12}$ s case, the data from the fitting curve give $\tau = 2.196$ ps and $v = 1.658$ cm$^2$/s. Comparison with the data used in the simulation ($\tau = 2.233$ ps and $v = 1.436$ cm$^2$/s) shows that this method gives quantitatively accurate $\tau$, and qualitatively accurate $v$ values. We also used the same method to extract values of $v$ and $\tau$ for other TeraFETs, and similar parameter accuracy was obtained.

Using the above method, we can also obtain the temperature variation of $v$, which is crucial for the characterization of TeraFET materials. In our simulation, the kinematic viscosity is obtained from equation (4) for degenerate condition ($T < T_F$). For non-degenerate case, the viscosity is fixed at $v(T = T_F)$ given by equation (4). For p-diamond, $T_F = 13.1$ K, thus the viscosity starts to increase after $T$ drops below 13.1 K. Fig. 13 illustrates both simulation and analytical results of $v$ versus $T$ at low temperature region. As seen, the calculated viscosities conform well to the analytical curve under the cryogenic temperature. This proves that the temperature variation of $v$ can be obtained from the voltage response signals of various temperatures.

IV. CONCLUSION

The hydrodynamic simulations show that the response of plasmonic TeraFETs to ultrashort pulse signals is strongly affected by the width of the incoming pulse ($t_{pw}$). In the short pulse mode ($t_{pw} \ll L/s$, where $L$ is the channel length, $s$ is the plasma velocity), the source-to-drain voltage response is a sharp pulse oscillatory decay preceded by a delay time on the order of $L/s$. The plasma wave travels along the channel like the shallow water wave waves with a relatively narrow wave package. In the long pulse mode ($t_{pw} > L/s$), the response profile has two oscillatory decay processes and the propagation of plasma wave is analogues to an oscillating rod with one side fixed. The ultimate response time at the long pulse mode is significantly larger than that under the short pulse conditions. The response time under the long pulse mode conforms well to the analytical theory for the step function response, since the detection conditions under the long pulse mode are close to the step response condition. The simulated waveform agrees well with the measured pulse response.

Our results could be used for the material parameter extraction from the pulse response data. The effective mass of the 2DEG carriers can be extracted from the delay time data, and the values of kinematic viscosity and momentum relaxation time can be obtained from the response time results under the long pulse mode regime. Using above method, it is also possible to extract the temperature dependence of the viscosity, which is very important for the characterization of 2DEG materials.